\documentclass[prb,floatfix,superscriptaddress,amsmath,amssymb,twocolumn,nobalancelastpage]{revtex4}

\usepackage{graphicx}
\usepackage{dcolumn}
\usepackage{bm}
\usepackage{pstricks}
\usepackage{datetime}
\usepackage{amsmath} 
\usepackage{float}
\usepackage{dsfont}

\def\nn{\nonumber}
\newcommand{\ba}{\begin{eqnarray}}
\newcommand{\ea}{\end{eqnarray}}
\def\be{\begin{equation}}
\def\ee{\end{equation}}

\def\bimno{Bi$_3$Mn$_4$O$_{12}$(NO$_3$)}

\begin{document}

\title{An effective field theory approach for the $S=3/2$ bilayer honeycomb antiferromagnet}

\author{S. Acevedo}
\email{santiago.acevedo@fisica.unlp.edu.ar}
\affiliation{IFLP - CONICET, Departamento de F\'isica, Universidad Nacional de La Plata,
C.C.\ 67, 1900 La Plata, Argentina.}

\author{ C. A.\ Lamas}
\email{lamas@fisica.unlp.edu.ar}
\affiliation{IFLP - CONICET, Departamento de F\'isica, Universidad Nacional de La Plata,
C.C.\ 67, 1900 La Plata, Argentina.}

\author{ P. Pujol}
\affiliation{Laboratoire de Physique Th\'eorique, CNRS and Universit\'e de Toulouse, UPS, Toulouse, F-31062, France.}

\begin{abstract}
The spin$-3/2$ Heisenberg antiferromagnet on the bilayer honeycomb  lattice is a minimal model to
describe the magnetic behavior of \bimno. We study this model with frustrating inter-layer second-neighbor couplings taking into account quantum and thermal 
fluctuations. We use a path integral formulation in terms of coherent states to describe the low energy physics of the model. 
We show that for a particular point in the parameter space, close to the experimental estimated couplings, a continuum classical degeneracy
is lifted by both quantum and thermal fluctuations, and a collinear state is then selected by an order by disorder mechanism.
Our results provide a global perspective in the understanding of the experimental observations.
\end{abstract}

\maketitle

\section{Introduction}
\label{sec:intro}

Frustrated magnetism is a prominent area of research with a broad range of sub fields,
 harboring new quantum states of condensed matter \cite{Moessner2006,Balents2010},
 topological ordering \cite{Wen1990,Wen2002,Kitaev2003} and addressing  candidates for quantum
computing \cite{Benjamin2015,Riste2015,Corcoles2015}.
%
Frustrated Heisenberg models on the Honeycomb lattice have become a paradigmatic example for the 
search of competing spiral order, lattice nematicity and plaquette valence bond
states \cite{Mulder,Okumura,Wang,Mosadeq,Cabra_honeycomb_prb, Ganesh_2011,
Albuquerque,Clark, Cabra_honeycomb_2, Mezzacapo,Bishop_2012,Li_2012_honeyJ1-J2-J3,
Bishop_2013,Fisher_2013, Ganesh_PRL_2013,Zhu_PRL_2013,Zhang_PRB_2013,Beca_2017}.
%
%
Furthermore, there are materials like the bismuth oxynitrate {\bimno}
 \cite{smirnova2009synthesis}, where the Mn$^{4+}$ ions of spin $3/2$ form honeycomb
 layers, with both nearest and next-nearest neighbor antiferromagnetic (AFM)
 exchange, and the Mn$^{4+}$ ions are grouped into pairs, so the resulting structure is a bilayer honeycomb lattice.
This compound has led research in bilayer honeycomb systems \cite{Ganesh_QMC,
Oitmaa_2012, Zhang2014, Arlego201415, Brenig2016, bishop2017frustrated,
Richter2017}. Most of these studies have focused on the stability of the
semi-classical phases, extending previous work on the single layer case.
The quantum phase diagram was partially studied for the $S=1/2$ case, emphasising the regions of the phase diagram where the ground state consists in a product of singlets, or quantum dimers, genuinely related to the bilayer geometry and not present in the single layer model\cite{Brenig2016}.
However, a complete understanding of the quantum phase diagram of the bilayer model is still missing.
\noindent
Inelastic neutron scattering measurements \cite{Matsuda2010} have been performed  in \bimno $\,$ at high magnetic fields, and more recently\cite{Matsuda2019} the magnetic couplings were estimated analyzing the magnetic dispersions using the linear spin-wave theory. The results for the couplings are consistent with previous results determined by ab initio density functional theory 
calculations \cite{Alaei2017}, suggesting that there is no significant frustration in the honeycomb plane but frustrating inter-layer interactions probably play an important role in destabilizing magnetic order.
The neutron scattering experiments show the presence of a short-range
antiferromagnetic order at low temperatures and the presence of a magnetic
transition in which the short-range order expands into a long-range Ne\'el order. 
The experimental data indicate that the collinear Ne\'el state becomes more stable at higher temperatures, i.e., thermal fluctuations stabilize the long-range Ne\'el order by an order by disorder mechanism\cite{Matsuda2010}.



Motivated by this experimental and ab initio results, in this article we study the honeycomb bilayer with in-plane first-neighbor  interactions and  competing inter-layer first- and second-neighbor interactions. 
In section \ref{sec:ground-state}, we start by introducing the model and recalling some results in the bilayer geometry where the ground state can be worked out exactly and exhibits a singlet product state. 
In section \ref{sec:field-theory} we use a path integral formulation \cite{Haldane, TTH} in terms of coherent states to describe the spin$-3/2$ model in the presence of a magnetic field.
We study both the weakly frustrated and strongly frustrated regimes. 
For the latter, we make the distinction between the dominant in-plane coupling case, and the dominant inter-layer coupling case. On one hand, section \ref{sec:dominant-in-plane-coupling} is relevant for \bimno, where we discuss how an order by disorder mechanism selects a staggered state as the one observed experimentally when the system is magnetized in the presence of an external magnetic field. Additionally, we argue how the system may exhibit magnetic short-range order for zero magnetization.
On the other hand, section \ref{sec:dominant-inter-layer-coupling} describes the formation of the valence bond solid studied in reference \onlinecite{Brenig2016} within the field theory approach. 

\section{Model and exact ground state }
\label{sec:ground-state}
%
%

\begin{figure}[t!]
\begin{center}
 \includegraphics[width=0.45\textwidth]{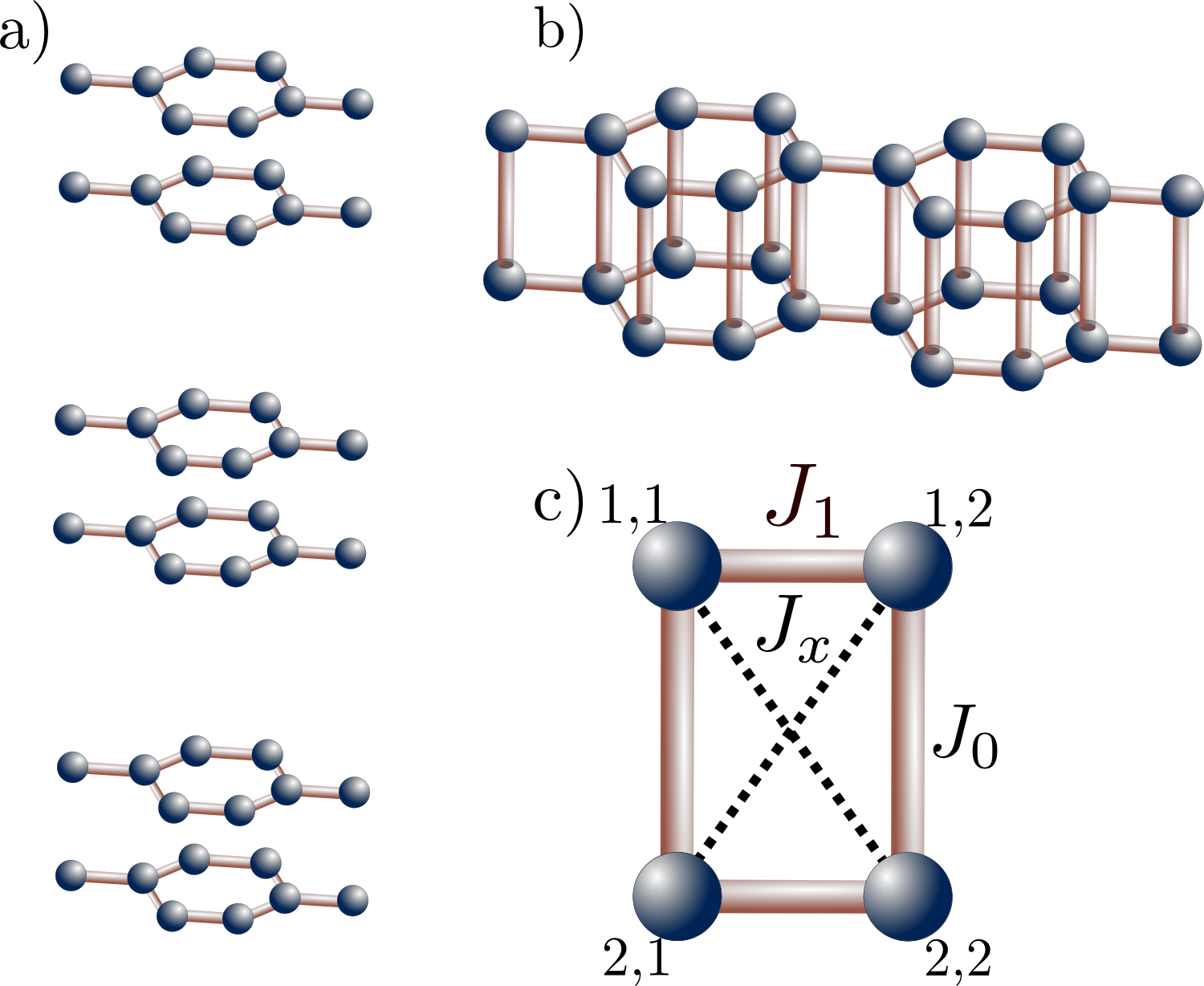}
\caption{\label{fig:bilayer1}The bilayer geometry considered here. Pannel a) shows the experimental structure corresponding 
to \bimno. b) Bilayer Honeycomb lattice corresponding to model
\eqref{eq:Hamiltoniano}. c) Four S=3/2 spins unit cell. The interactions in the Hamiltonian correspond to the in-plane 
nearest neighbor coupling $J_{1}$, the inter-layer nearest neighbor coupling $J_{0}$ and the inter-layer next-nearest neighbor coupling $J_{x}$. }
\end{center}
\end{figure}

The model of interest consists of quantum spins$-3/2$ on a bilayer honeycomb lattice, depicted in figure (\ref{fig:bilayer1}).
The spins are coupled through Heisenberg isotropic interactions $J_{1}$, $J_{x}$ and $J_{0}$. 
For \bimno, experimental and numeric results\cite{Matsuda2019} show that $J_1$ is the dominant coupling, and $J_0$ is roughly  three times the 
value of the $J_x$. 

\noindent
To connect with previous results in this system\cite{Brenig2016}, let us write the Hamiltonian as a sum of terms over the square plaquettes schematized in figure (\ref{fig:bilayer1}-c).
\begin{eqnarray}
\lefteqn{
H=\sum_{\boldsymbol{r}}\sum_{i=0}^2\bigg\{
\frac{J_{0}}{3}\left( \mathbf{S}_{1,1}(\boldsymbol{r})\cdot
\mathbf{S}_{2,1}(\boldsymbol{r})
+\mathbf{S}_{1,2}(\boldsymbol{r})\cdot
\mathbf{S}_{2,2}(\boldsymbol{r})\right)
\nonumber}
\\
&+& J_{1}\left(
\mathbf{S}_{1,1}(\boldsymbol{r}_i)\cdot \mathbf{S}_{1,2}(\boldsymbol{r})+
\mathbf{S}_{2,1}(\boldsymbol{r}_i)\cdot \mathbf{S}_{2,2}(\boldsymbol{r})
\right)\hphantom{aaaaaa}\nonumber\\
&+&  J_{x}\left(
\mathbf{S}_{1,1}(\boldsymbol{r}_i)\cdot \mathbf{S}_{2,2}(\boldsymbol{r})+
\mathbf{S}_{2,1}(\boldsymbol{r}_i)\cdot \mathbf{S}_{1,2}(\boldsymbol{r})
\right)\bigg \} \\
&-& h \sum_{\boldsymbol{r}} \bigg \{ S^z_{1,1}(\boldsymbol{r}) + S^z_{1,2}(\boldsymbol{r}) + S^z_{2,1}(\boldsymbol{r}) + S^z_{2,2}(\boldsymbol{r})\bigg \}
\label{eq:Hamiltoniano}
\end{eqnarray}

\noindent
in which the $z$-axis is oriented in the direction\footnote{In the absence of magnetic field the Hamiltonian is full $SU(2)$ symmetric. Thus, the direction of the magnetic field applied here can be arbitrary.} of the external  magnetic field h,
$i=0,1,2$ corresponds to $\boldsymbol{r}_{ (0,1,2) }= \boldsymbol{r} +(\boldsymbol{0}
,\boldsymbol{e}_{1} ,\boldsymbol{e}_{2})$, being $\boldsymbol{e}_{1}$ and $\boldsymbol{e}_{2}$ the
primitive vectors of the triangular lattice. 

When studying the low-energy theory of this model we will see that for some particular values of the couplings we 
find points where the theory presents zero modes. One of these points corresponds to the 
case $J_{1}=J_{x}$ and $J_{0}$ large, where the ground state can be determined exactly and consists in a product of singlets.
In order to see how the fine tuning works, it is useful to introduce the bond spin operators
\begin{equation}
\boldsymbol{L}_{\eta}=\boldsymbol{S}_{1,\eta}+\mathbf{S}_{2,\eta}
\hspace{1cm}
\boldsymbol{K}_{\eta}=\mathbf{S}_{1,\eta}-\mathbf{S}_{2,\eta},
\label{eq:L-K}
\end{equation}
with $\eta=1,2$, where $[L_\eta^\alpha,L_{\eta'}^\beta]=i \epsilon^{\alpha \beta \gamma} L_\eta^\gamma \delta_{\eta,\eta'}$, $[L_\eta^\alpha,K_{\eta'}^\beta]=i \epsilon^{\alpha \beta \gamma} K_\eta^\gamma \delta_{\eta,\eta'}$ and $[K_\eta^\alpha,K_{\eta'}^\beta]=i \epsilon^{\alpha \beta \gamma} L_\eta^\gamma \delta_{\eta,\eta'}$ in the same unit cell $\boldsymbol{r}${, being $\epsilon^{ \alpha \beta \gamma}$ the fully anti-symmetric Levi-Civita tensor and $\epsilon^{xyz}=1.$}

The Hamiltonian written in terms of the bond operators is 

\begin{equation}
\begin{split}
 H=&-2NJ_0 S(S+1) + 
 \frac{1}{2} \sum_{\boldsymbol{r},i} \bigg \{  \frac{J_0}{3} 
 \bigg (\boldsymbol{L}_{1}^2(\boldsymbol{r}_i) + \boldsymbol{L}_2^2(\boldsymbol{r}) \bigg )  + \\
 & (J_1+J_x) \; \,\boldsymbol{L}_1(\boldsymbol{r}_i) \cdot \boldsymbol{L}_2(\boldsymbol{r}) + 
 (J_1-J_x) \boldsymbol{K}_1(\boldsymbol{r}_i )\cdot \boldsymbol{K}_2(\boldsymbol{r})
 \bigg \} \\
 & - h \sum_{\boldsymbol{r}} \bigg \{  \boldsymbol{L}_1^z(\boldsymbol{r}) + \boldsymbol{L}_2^z(\boldsymbol{r})
 \bigg \},
\end{split}
\label{eq:H0}
\end{equation}
with $N$ the number of unit cells in the system and $S$ the spin quantum number. 

From Eq (\ref{eq:H0}) we can see clearly that for $J_{1}=J_{x}$, the last term in
 the Hamiltonian vanishes and the Hamiltonian depends only on the bond spin
 $\boldsymbol{L}_{\eta}(\boldsymbol{r})$. Therefore, at $J_{1}=J_{x}$, the eigenstates of $H$ are
 multiplets of the total bond spin. Among those is the {\em product state of
 bond singlets}, i.e.
 $|\psi\rangle = \bigotimes_{\boldsymbol{r}} 
 | s_{1} \rangle_{ \boldsymbol{r}} 
 |s_{2} \rangle_{\boldsymbol{r}}$ with 
 $\boldsymbol{L}_{\eta}(\boldsymbol{r})
 |s_{\eta}\rangle_{\boldsymbol{r}}=0$, and 
 $|s_{\eta} \rangle_{\boldsymbol{r}} = \sum_{m =-S}^S (-1)^{S-m} |m,-m\rangle /\sqrt{2 S+1}$. 
 Here $|m,-m\rangle$ labels a product of eigenstates of $\boldsymbol{S}^{z}_{1,\eta}(\boldsymbol{r})$ and
 $\boldsymbol{S}^{z}_{2,\eta}(\boldsymbol{r})$ on
 dimer $\eta$ of the unit cell located at $\boldsymbol{r}$.

The preceding is valid for {\em any} site spin  $S$, and for $J_0 \gg J_1=J_x$ the valence bond solid described before is the ground state of the system both in the absence of a magnetic field or with an small magnetic field compared to the resulting magnetic gap. If we increase the value of $J_1=J_x$ there will be a phase transition at some $J_1^*$ where the nature of the ground state, as well as the value of $J_1^*$ should depend on the spin $S$.
Below we show that at the semi-classical level, the fine tuning $J_1=J_x$ gives rise to a zero mode in the effective theory.

%


\begin{section}{Semiclassical effective field theory}
\label{sec:field-theory}

We write an effective field theory for the system using the coherent-state path integral description developped by Haldane\cite{Haldane} and Tanaka et al\cite{TTH}. The spins are represented by $O(3)$ vectors of modulus $S$: 
$\boldsymbol{S}=S (\cos{\phi}\sin{\theta}, \sin{\phi}\sin{\theta},\cos{\theta})$, and we consider quantum fluctuations on top of a lowest-energy configuration of the classical system. 
Our parametrization for the classical lowest-energy configuration in the presence of a magnetic field $h$ consists in a canted N\'eel configuration, where for each unit cell we set 
\begin{equation}
 \begin{split}
  \phi^0_{l,\eta}(\boldsymbol{r})&= \pi (l+\eta)\\
  \theta^0_{l,\eta}(\boldsymbol{r})&= \theta^0(h,J_0,J_1),
 \end{split}
 \label{eq:parametrizacion0}
\end{equation}
for $l=1,2$ and $\eta=1,2$.
This configuration corresponds to the low-frustration limit $J_x \ll J_1,J_0$. The classical energy at $T=0$ is then minimized by $\cos{\theta^0}=h/2(J_0+3J_1)$.
In the continuum limit quantum fluctuations are then added over the classical ground state changing $\phi^0_{l,\eta}(x,y) \rightarrow \pi(l+\eta) + \phi_{l,\eta}(x,y)$ and 
$\theta^0_{l,\eta}(x,y) \rightarrow \theta^0 + \delta \theta_{l,\eta}(x,y)$. 
The canonical conjugate fields of the theory are $\phi_{l,\eta}$ and $a\Pi_{l,\eta}=-S \big ( \delta \theta \sin{\theta^0} + \frac{1}{2}(\delta \theta)^2 \cos{\theta_0} \big )$, where $a$ is the distance between the spins connected by $J_1$ in the unit cell. The spin operators written to quadratic order in $\phi_{l,\eta}$ and $\Pi_{l,\eta}$ are

\begin{equation}
\begin{split}
  S^z_{l,\eta}&= S \cos{\theta_0} + a\Pi_{l,\eta} \\
S^{\pm}_{l,\eta}&=(-1)^{l+\eta} e^{\pm i \phi_{l,\eta}} \bigg (
S \sin{\theta^0}- \frac{m}{S \sin{\theta^0}} a\Pi_{l,\eta} \\-
&\frac{1}{2} \frac{S^2}{S^2-m^2} \frac{1}{S \sin{\theta^0}} (a \Pi_{l,\eta})^2
\bigg ),
\end{split}
\label{eq:classical-spins}
\end{equation}
which fulfill the SU(2) algebra replacing the quantum commutator by a classical Poisson bracket. In the last equation we defined $m=S \cos{\theta^0}$.
Replacing \eqref{eq:classical-spins} in the Hamiltonian and keeping terms up to quadratic order, we obtain the non-interacting effective action $\mathcal{S}$ given by
\begin{equation}
 \mathcal{S}=\mathcal{S}_{cl}+\mathcal{S}_{BP},
 \label{eq:total_action_def}
\end{equation}
where $\mathcal{S}_{cl}$ is the classical action of the system, which contains kinetic terms and mass terms, i.e.,
$\mathcal{S}_{cl}=\mathcal{S}_K + \mathcal{S}_M$. In order to simplify the notation using a single index for the fields, we rename the fields as 
\begin{equation}
\nn
 \phi_{i,j}\rightarrow \phi_{i(i-1)+j},
\end{equation}
obtaining for the action the following expressions 
\begin{equation}
\begin{split}
  \mathcal{S}_K&=  \int d\tau \frac{d^2x}{\nu} \frac{K}{2a^2}
 \sum_{j=1}^2 \bigg (( \boldsymbol{e}_j \cdot \boldsymbol{\nabla} \phi_{1} ) ^2 + ( \boldsymbol{e}_j \cdot\boldsymbol{\nabla} \phi_{3})^2   \bigg)\\ 
 &=\int d\tau \frac{d^2x}{\nu} \frac{K}{2 }   \frac{3 }{2}\bigg ( 
 3(\partial_x \phi_1 )^2 + 3(\partial_x \phi_3 )^2  \\ 
 &\qquad \qquad \qquad \qquad   +(\partial_y \phi_1 )^2 +  (\partial_y \phi_3 )^2
 \bigg ),
\end{split}
\end{equation}
with $\tau \in [0, \beta]$ the imaginary time, $\beta$ the inverse temperature (where Boltzmann constant is absorbed),  $\nu=(9\sqrt{3} a^2)/2$, $K=(Sa)^2  (J_1-J_x) \sin^2{(\theta^0)}$, $\boldsymbol{e}_j=a \frac{\sqrt{3}}{2} (\sqrt{3},(-1)^{j+1}) $, $j=1,2$, and

\begin{equation}
 \mathcal{S}_M= \int d\tau \frac{d^2x}{\nu} \bigg ( \frac{1}{2} \phi_i (M_\phi)_{ij} \phi_j + 
 \frac{a^2}{2} \Pi_i (M_\Pi)_{ij} \Pi_j
  \bigg ),
\end{equation}
where Einstein's notation is being used for the repeated indices, and $M_\phi$ and $M_\Pi$ are symmetric mass matrices.
The second term in \eqref{eq:total_action_def} is the Berry phase term, which arises from the non-orthogonality of the coherent-state basis, and is given by 
\begin{equation}
\begin{split}
 \mathcal{S}_{BP}= -i (S-m) &\int d\tau \frac{d^2x}{\nu} \sum_{j} \partial_\tau \phi_{j}  \\
  + i & \int d\tau \frac{d^2x}{\nu} \sum_{j} (\partial_\tau \phi_{j}) a\Pi_{j}.
\end{split}
\end{equation}
The symmetric mass matrix $M_\phi$ is diagonalized by the transformation 
\begin{equation}
 \text{W=} \frac{1}{4}\left(
\begin{array}{cccc}
 1 & 1 & 1 & 1 \\
 1 & -1 & -1 & 1 \\
 -1 & -1 & 1 & 1 \\
 -1 & 1 & -1 & 1 \\
\end{array}
\right),
\label{eq:transformation}
\end{equation}
where $\phi'_j=\sum_{k=1}^4 W_{jk}\phi_k$, $j=1,...,4$ and $M'_\phi=4W M_\phi W^{-1} =\text{diag}(m_1,m_2,m_3,m_4)$, with
\begin{equation}
 \begin{cases}
 m_1=0 \\
  m_2=8 S^2 \sin ^2(\theta^0) (J_0+3 J_1)\\
  m_3=8 S^2 \sin ^2(\theta^0) (J_0-3J_x)\\
  m_4=24 S^2\sin ^2(\theta^0) (J_1 - J_x).
 \end{cases}
 \label{eq:m}
\end{equation}
The nullity of the first mass is know to prevail to all order in the development and reflects the protection of the $\phi'_1$ field which is the Goldstone field associated to the $U(1)$ symmetry of the Hamiltonian.
The same transformation diagonalices $M_\Pi$ as well, i.e., $M'_\Pi=4W M_\Pi W^{-1} =\text{diag}(\mu_1,\mu_2,\mu_3,\mu_4)$ with

\begin{equation}
 \begin{cases}
  \mu_1= 8 (J_0+3 J_1) \\
  \mu_2=8 (J_0+3 J_1) \cot^2{(\theta^0)} \\
  \mu_3=8 \big( -J_0+3 J_1+(J_0-3 J_x) \csc^2{(\theta^0)} \big)\\
  \mu_4=8 \big(J_0-3 J_1+3 (J_1-J_x) \csc^2{(\theta^0)}\big).
 \end{cases}
 \label{eq:mu}
\end{equation}
The complete effective action after the transformation $W$ takes the form
\begin{equation}
\begin{split}
 \mathcal{S}=\int d\tau \frac{d^2x}{\nu}& \bigg \{
 \frac{K}{2} 3  \bigg(  
  3(\partial_x \phi_2' - \partial_x \phi_3' )^2+
  3(\partial_x \phi_1' - \partial_x \phi_4' )^2 +\\
 &(\partial_y \phi_2' - \partial_y \phi_3' )^2 +
 (\partial_y \phi_1' - \partial_y \phi_4' )^2
 \bigg)+ \\
 &\frac{1}{2} \sum_{j=2}^4 \bigg ( m_j  \phi_j'^2 \bigg ) +
 \frac{1}{2} \sum_{j=1}^4  \bigg( \mu_j   (a\Pi_j')^2  \bigg ) +\\
 & (-4i) (S-m) (\partial_\tau \phi'_1) + 4i \sum_{j=1}^4 (\partial_\tau \phi'_j) a \Pi'_j
 \bigg \}.
\end{split}
\end{equation}

In general all mass terms of the $\Pi'_j$ fields are non-zero, and as so these fields are short-ranged.  Since they are not expected to contribute at large scales, we can integrate them out, obtaining the effective action 

\begin{equation}
\begin{split}
  \mathcal{S}=\int d\tau \frac{d^2x}{\nu} &\bigg \{
 \frac{K}{2} 3\bigg(  
  3(\partial_x \phi_2' - \partial_x \phi_3' )^2+
  3(\partial_x \phi_1' - \partial_x \phi_4' )^2+\\
 &(\partial_y \phi_2' - \partial_y \phi_3' )^2 +
 (\partial_y \phi_1' - \partial_y \phi_4' )^2
 \bigg)+ \\
  &\frac{1}{2} \sum_{j=2}^4 \bigg ( m_j \phi_j'^2 \bigg ) + \frac{1}{2}
\sum_{j=1}^4 \bigg ( \frac{16}{\mu_j} (\partial_\tau \phi_j')^2 \bigg ) +\\
&(-4i)(S-m) (\partial_\tau\phi'_1)
 \bigg \}.
\end{split}
\label{eq:full-action}
\end{equation}

In the following, we will discuss some important regimes of this theory.

\subsection{Weakly frustrated regime}

For $J_x \ll J_0,J_1$ the system is weakly frustrated and we have a non-vanishing $m_i$, $\mu_j$, $i=2,...,4$, $j=1,...,4$. 
As mentioned before, the field $\phi'_1$ has no mass term because of the $U(1)$ symmetry, and in fact remains long-ranged for $4(S-m) \not\in \mathds{Z}$ and $T=0$. 

\noindent
The factor $1/4$ in the transformation \eqref{eq:transformation} is chosen to give the field $\phi_1'$ the correct periodicity\cite{Lamas2011}. If the fields $\phi'_2$, $\phi'_3$ and $\phi_4'$ are gapped, they can be set to zero in the low energy limit, meaning $\phi_1=\phi_2=\phi_3=\phi_4$. In this case the action depends only on the gappless field $\phi_1'=\phi_1$. 
Thus, the action \eqref{eq:full-action} corresponds to a classical $XY$ model with an additional Berry phase term. If $T=0$, the theory is in three dimensions, whereas for finite temperature $ 0 \leq \tau \leq \beta < \infty$, and the theory is two-dimensional, where Mermin-Wagner theorem forbids long-range order.

\noindent
The condition $4(S-m) \not\in \mathds{Z}$ corresponds to the presence of a non-trivial Berry phase. For a spin in position $(x_0,y_0)$ that has a phase winding of $2 \pi$ when evolving in imaginary time there is a contribution to the effective action of $\delta \mathcal{S}=(-4 i)(S-m) 2\pi$. The Berry phase forbids vortices to contribute to the partition function because they enter in the partition function weighted with an oscillatory phase factor which leads to destructive interference\cite{TTH}. 
For $T=0$, these vortices are the only mechanism available to destroy the Long-Range Antiferromagnetic Order (LRAFO) for the in-plane spin 
components represented by the $\phi'_1$ field. As vortices are forbidden for a generic value of the magnetization, the LRAFO is preserved at $T=0$. For non-zero $T$, the LRAFO becomes a Quasi-Long-Range Antiferromagnetic Order (QLRAFO). 

\noindent
If $4(S-m)$ is an integer the Berry phase term trivializes and can be dropped. This condition is known as the OYA\cite{OYA} (Oshikawa-Yamanaka-Affleck) criterium, and is a well known necessary condition for a magnetic plateau to emerge. Vortices of $\phi'_1$ are able to proliferate if that is energetically favourable (i.e. if the stiffness is small enough), disordering the field $\phi'_1$ and opening a gap in the system. The resulting phase has short-range antiferromagnetic order. Finally, the special case where $4(S-m)$ is a rational number can give rise to more exotic phases, as developed in Ref. \onlinecite{TTH}.

\begin{figure}[t!]
\begin{center}
 \includegraphics[width=0.45\textwidth]{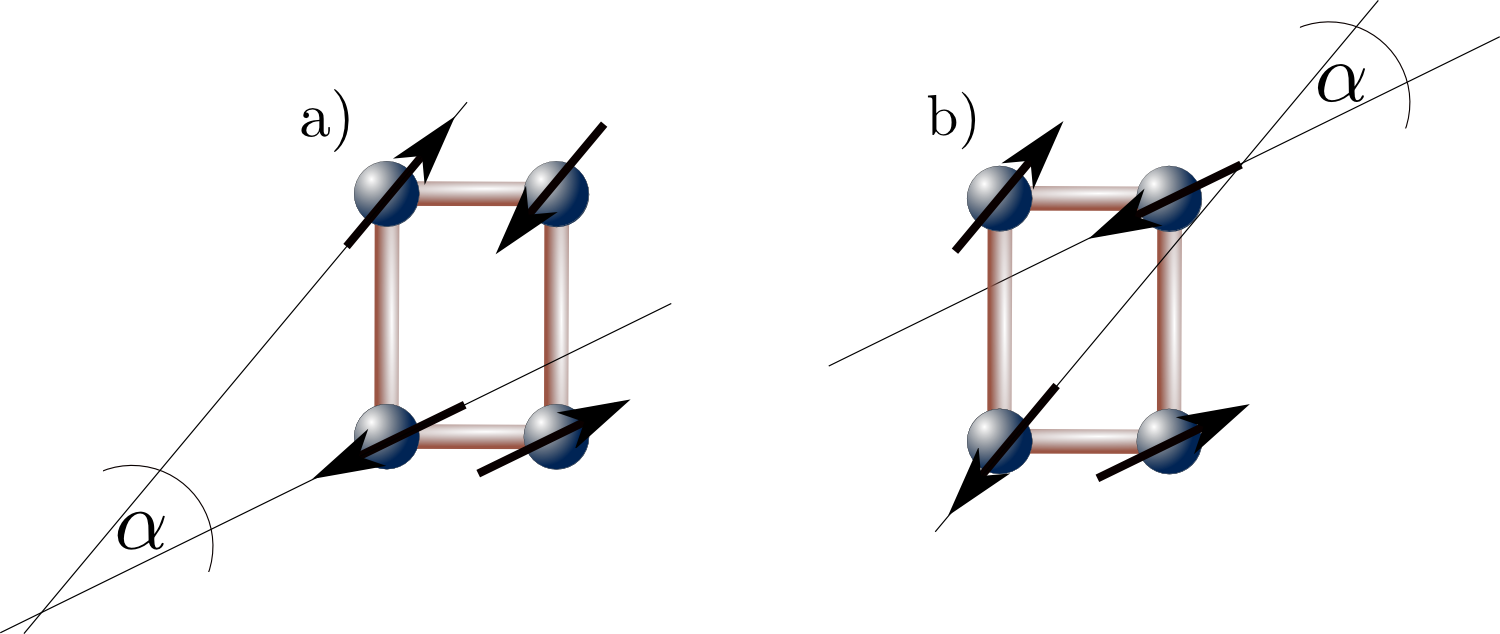}
\caption{\label{fig:zero-modes}Schematic representation of the zero modes corresponding to a) $J_{x}=J_{0}/3$ and b) $J_{1}=J_{x}$. }
\end{center}
\end{figure}

\subsection{Frustrated regime}

The situation described above changes when the values of the couplings promote competition between terms of the Hamiltonian. 
We discuss below those situations in two limits, where the dominant interaction is the in-plane or the inter-layer coupling.

\subsubsection{Dominant in-plane coupling}
\label{sec:dominant-in-plane-coupling}

If $J_1$ is the dominant coupling (as expected for \bimno)\cite{Matsuda2019}, increasing $J_x$ we find that $m_3=0$ at $J_x=J_0/3$ (see \eqref{eq:m}), 
whereas $\mu_j>0,$ $j=1,...,4$.
It is easy to see that the condition $J_x=J_0/3$ leads not only to a vanishing mass in the effective theory but it corresponds to a zero mode of the Hamiltonian using the parametization \eqref{eq:parametrizacion0}. 
Indeed, if we take the Hamiltonian \eqref{eq:Hamiltoniano} evaluated in $J_x=J_0/3$,  we can make a variation in the classical parametrization \eqref{eq:parametrizacion0} introducing the real parameters $\alpha_{l,\eta}$ by
\begin{equation}
 \phi^{0}_{l,\eta} =(l+\eta) \pi + \alpha_{l,\eta},
 \label{eq:parametrizacion-alpha}
\end{equation}
where $l=1,2$, $\eta=1,2$. If we take $\alpha_{l,1}=\alpha_{l,2}$, $l=1,2$, then the Hamiltonian is a function of $\theta^0$ only. As a consequence of the $U(1)$ symmetry of the system we can set 
$\alpha_{2,1}=0$, leaving only one free parameter $\alpha_{1,1}\equiv \alpha$.
The parameter $\alpha$ corresponds to a relative angle between spins in the two hexagonal layers as we show in Fig. \ref{fig:zero-modes}-a), and 
enters in the effective field theory \textit{only} through the kinetic term, 
since it is a zero mode of the theory for fluctuations that are uniform in space. 

Here we make the low-energy approximation $\phi'_2=\phi'_4=0$ for the massive fields, which implies
$\phi_2=\phi_1 $ and $\phi_4=\phi_3$, 
and we study the massless theory 
\begin{equation}
\begin{split}
  \mathcal{S}=\sqrt{3}\int d\tau d^2x \bigg \{
 \frac{\tilde{K}}{2}\bigg(  
  (\boldsymbol{\nabla}\phi_a )^2+
  (\boldsymbol{\nabla} \phi_s  )^2
 \bigg)+ \\
\frac{1}{2\tilde{\mu}_s} (\partial_\tau \phi_s)^2+ \frac{1}{2\tilde{\mu}_a} (\partial_\tau \phi_a)^2 + (-4i)\frac{(S-m)}{\nu} (\partial_\tau\phi_s)
 \bigg \},
\end{split}
\end{equation}
where $\tilde{K}=\frac{3}{\nu}(Sa)^2(J_1-J_x \cos{\alpha})\sin^2{\theta^0}$,
$\tilde{\mu}_{s,a}=\frac{\nu}{16} \mu_{s,a}$, $\phi_s=(\phi_1+\phi_3)/2$,
$\phi_a=(-\phi_1+\phi_3)/2$. 
Notice that all the $\alpha$-dependence is in the parameter $\tilde{K}$ and we have made a rescaling $x \rightarrow x/3$. 

In the following we drop the factor $\sqrt{3}$ outside the integral in the last equation.
The notation $\phi_s$ and $\phi_a$ is chosen to emphasize the presence of a symmetrical and an antisymmetrical combination of the fields.
At this order the partition function is factorized as
\begin{equation}
\nn
Z=\bigg ( \int \mathcal{D} \phi_s e^{-\mathcal{S}[\phi_s]} \bigg ) \bigg (\int \mathcal{D} \phi_a e^{-\mathcal{S}[\phi_a]} \bigg )=Z_s Z_a,
\end{equation}
and can be computed analytically.
Here we follow the approach from Ref. \onlinecite{LCPR}, and we express $\phi_b$, $b \in \{ s,a \}$ in terms of the crystalline momentum $k$ and the Matsubara 
frequencies $\omega_n=\frac{2 \pi n}{\beta}$, $n \in \mathds{Z}$, by
\begin{equation}
 \phi_{b}(\boldsymbol{r},\tau)=
 \sum_{n=-\infty}^\infty \frac{1}{2\pi \beta} \int d^2 k e^{i \boldsymbol{k} \boldsymbol{r}} e^{- i \omega_n \tau} \phi_b(\boldsymbol{k},\omega_n).
 \label{eq:Matsubara}
\end{equation}
This representation is valid only for fields satisfying $\phi_b(\boldsymbol{r},0)=\phi_b(\boldsymbol{r},\beta)$, i.e.,
without vorticity, and yields $S_{BP}=0$.


The complete action reads
\begin{equation}
\nn
 S=\frac{1}{\beta} \sum_b \sum_{n=-\infty}^\infty  \frac{1}{2} \int d^2k  \phi_b^* (\boldsymbol{k},\omega_n)\phi_b(\boldsymbol{k},\omega_n) 
 \bigg ( \tilde{K} k^2 + \frac{\omega_n^2}{\tilde{\mu}_b} \bigg ).
\end{equation}
We calculate the Gaussian functional integral in terms of a dimensionless rescaled field
$\phi'(k,\omega_n) = \phi(k,\omega_n) \Gamma/\beta$, where $\Gamma \propto 1/a^2$, obtaining
\begin{equation}
\nn
 \log{(Z_b)} = \frac{-1}{2} \int \frac{d^2k}{\Gamma} \sum_{n=-\infty}^\infty
 \log{\bigg [ \frac{\beta}{\Gamma} \bigg(\tilde{K} k^2+\frac{\omega_n^2}{\tilde{\mu}_b}\bigg) \bigg ]} .
\end{equation}
To sum the series, we make use of the identity
\begin{equation}
\nn
\begin{split}
 \int_{1}^{\beta \tilde{K} k^2 /\Gamma} \frac{dt^2}{\frac{\omega_n^2}{\tilde{\mu}_b \Gamma /\beta}+ t^2}=&
 \log{\bigg(\frac{\beta}{\Gamma}\tilde{K}  k^2+\frac{\omega_n^2}{\tilde{\mu}_b \Gamma/\beta}  \bigg)}\\
 -& \log{\bigg(1+\frac{\omega_n^2}{\tilde{\mu}_b \Gamma/\beta}\bigg)}.
 \end{split}
\end{equation}
Hence, the Helmholtz free energy is
\begin{equation}
\nn
 F=-\frac{1}{\beta} \log{(Z)}=\frac{1}{\beta} \int \frac{d^2 k}{\Gamma} \sum_b
 \log \bigg( \sinh{\bigg(\frac{1}{2} k \beta \sqrt{\tilde{K} \tilde{\mu}_b} \bigg)} \bigg ),
\end{equation}
where we have dropped vacuum contributions, i.e., terms independent of the momentum $k$ and the angle $\alpha$. The last equation can be rewritten as
\begin{equation}
\nn
 F=\int \frac{d^2 k}{\Gamma}\sum_b \bigg \{ \frac{1}{2}  k \sqrt{\tilde{K} \tilde{\mu}_b}  +
 \frac{1}{\beta} \log{\bigg( 1- e^{-k \beta \sqrt{{\tilde{K} \tilde{\mu}_b}}} \bigg)}
 \bigg \},
\end{equation}
where first term is the quantum contribution to the free energy, $F_Q$, and the second one is the thermal contribution, $F_\beta$.
The former is integrated directly, giving

\begin{figure}[t!]
\begin{center}
 \includegraphics[width=0.48\textwidth]{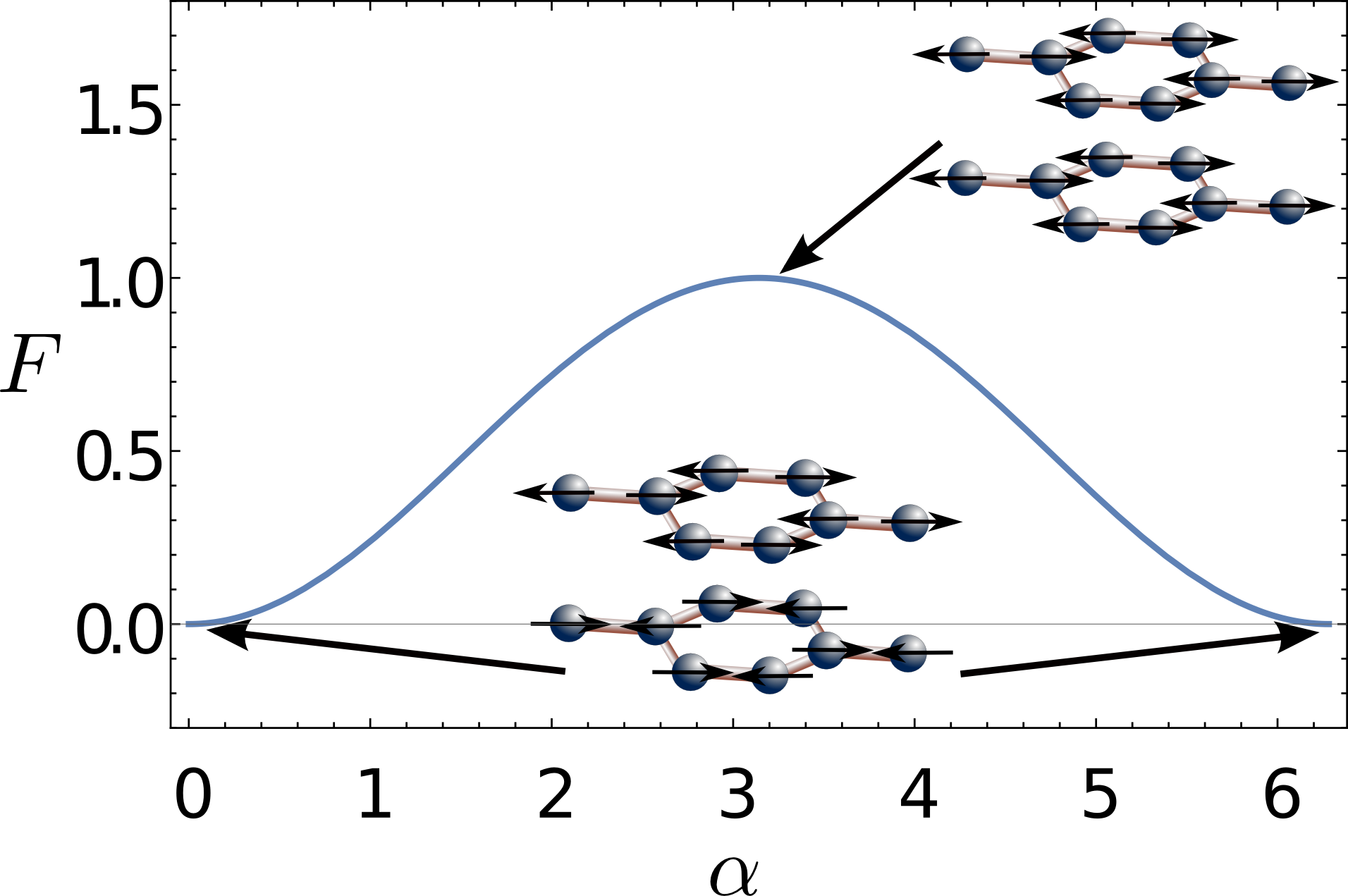}
\caption{\label{fig:free-energy} Free Energy corresponding to Eq. (\ref{eq:Ffinal}),
with $J_1=1$, $J_x=1/10$, $J_0=3 J_x$, $\theta^0=\pi/4$ and $\beta=5$. The Free energy presents  minimums at $\alpha=0,2\pi$ corresponding
to antiparallel configuration in the vertical bonds. The maximum at $\alpha=\pi$ corresponds to a parrallel configuration in the vertical bonds.
For the presentation, have normalized the free energy such that $F(\alpha=0)=0$ and the maximum value of $F$ is 1.\\
}
\end{center}
\end{figure}

\begin{equation}
\nn
 F_Q=\frac{\pi}{3 \Gamma} \sqrt{\tilde{K}} \Lambda^3 \sum_b \sqrt{\tilde{\mu}_b},
\end{equation}
where $\Lambda=2 \pi/a$ is the momentum cut-off. For $F_\beta$ we have
\begin{equation}
\nn
 F_\beta=\frac{2\pi}{\beta^3 \Gamma \tilde{K}} \sum_b \frac{1}{\tilde{\mu}_b} 
 \int_0^{\Lambda \beta \sqrt{\tilde{K} \tilde{\mu}_b}} dx  \; x  \log \bigg(
 1-e^{-x}
 \bigg),
\end{equation}
where we define the dimensionless variable $x=k\beta \sqrt{\tilde{K} \tilde{\mu}_b}$. 
Here we take the low temperature limit $\beta \Lambda \sqrt{\tilde{K} \tilde{\mu}_b} \gg 1$ and we obtain
\begin{equation}
\nn
 F_\beta=(-\zeta(3)) \frac{2\pi}{\beta^3 \Gamma \tilde{K}} \sum_b \frac{1}{\tilde{\mu}_b},
\end{equation}
where $-\zeta(3) \equiv \int_0^\infty dx \; x \log \bigg( 1-e^{-x}\bigg)\approx -1.2$, being $\zeta(s)$ the Riemann zeta function, for $Re(s)>1$.

The complete Gaussian (non-interacting) free energy in terms of the bare microscopic parameters and the angle $\alpha$ is
\begin{widetext}
\begin{equation}
 F= 6 \sqrt{2} \pi ^4 \left(\sqrt{J_1-J_x}+\sqrt{J_1+J_x}\right)\sin{(\theta^0 )} \sqrt{ J_1- J_x \cos (\alpha )} 
-\frac{32 \pi  J_1 \zeta(3) \csc ^2(\theta^0 )}{81 \beta ^3 \left(J_1^2-J_x^2\right) (J_1- J_x \cos (\alpha ))}.
\label{eq:Ffinal}
\end{equation}
\end{widetext}

Both the thermal and the quantum contribution to the free energy have a single minimum at $\alpha=0$, so the N\'eel configuration is selected by both thermal and quantum fluctuations. 
Additionally, we see in \eqref{eq:Ffinal} that in this approximation, for $J_1=J_x$ the quantum contribution vanish and the thermal contribution diverges, suggesting that near $J_1=J_x=J_0/3$ the thermal contributions dominate against the quantum ones. A classical analysis at finite temperature was done for this system with $J_1=J_x=J_0/3$ in Ref. \onlinecite{FlaviaDiegoMC}. For this particular choice of magnetic couplings our effective field theory is unstable, given that $m_3=m_4=\mu_3=\mu_4=0$. Nonetheless, the presence of a anisotropy term $H_D=\sum_{\boldsymbol{r},l,\eta} D (S^z_{l,\eta})^2$, with $D>0$,  would provide a finite contribution to all four  $\mu_j$, $j=1,...,4$, stabilizing the magnetization fluctuations. This scenario is beyond the scope of our present work because the point $J_1=J_x=J_0/3$ is far away from the experimental and numerical estimations of the magnetic interactions in \bimno, where $J_1$ is the dominant coupling\cite{Matsuda2019}. Besides, our classical ground state corresponds to the limit of low frustration whereas for $J_1=J_x=J_0/3$ frustration is dominant and leads to more general classical ground states.\cite{FlaviaDiegoMC}

So far the results in this subsection apply for vorticity-free fields $\phi_s,$ $\phi_a$. If instead vortices of $\phi_s$ proliferate the system enters a short-ranged gapped phase. For this to be allowed the theory must not have a Berry phase term, i.e., the condition $4(S-m) \in \mathds{Z}$ must hold. In this scenario the symmetric field $\phi_s$ disorders and consequently the field $\phi_a$ disorders as well, because both combinations of fields are not independent.

\subsubsection{Dominant inter-layer coupling}
\label{sec:dominant-inter-layer-coupling}
If instead in \eqref{eq:full-action} $J_0$ is the dominant coupling and we increase the magnitude of $J_x$, the theory becomes singular at $J_1=J_x$. 
In this case both the mass $m_4$ and the stiffness $K$ simultaneously vanish. In contrast $\mu_j$ remains positive for $j=1,...,4$ (see \eqref{eq:mu}). 
As before, the condition $J_1=J_x$ corresponds to  a zero mode of the Hamiltonian parametrized by \eqref{eq:parametrizacion0}. In this case we can again make a variation in the parametrization by adding the real parameters $\alpha_{l,\eta}$, $l=1,2$, $\eta=1,2$ from equation \eqref{eq:parametrizacion-alpha}.
If we now set $\alpha_{1,\eta}=\alpha_{2,\eta}$, $\eta=1,2$, the classical energy at $T=0$ is a function of $\theta^0$ only. Again, because of the $U(1)$ symmetry of the system we can take $\alpha_{2,1}\equiv \alpha$ and $\alpha_{1,2}=0$. This zero mode is presented graphically in Fig. \ref{fig:zero-modes}-b). The parameter $\alpha$ does not enter in the effective theory this time because the kinetic term vanishes. 

For $J_1=J_x$, in the low-energy limit the two remaining massive fields $\phi'_2$ and $\phi'_3$ in \eqref{eq:full-action} may be set to zero, meaning $\phi_1=\phi_3$ and $\phi_2=\phi_4$.
If we consider the case $m=0$ then there is no Berry phase and the vortices in $\phi'_1$ may proliferate if energetically favourable, leading the system to a gapped short-range phase. In this scenario the gapless field $\phi'_4$ gets delocalized, and its canonical-conjugate field $\Pi'_4$ gets localized. The vanishing of the stiffness produces a flat band in the dispersion relation of the magnetic excitations, which imply that they are localized in coordinate-space, i.e., do not propagate through the lattice. Similar magnon-crystal ground states are present in other frustrated quantum spin system in one-, two-, and three-dimensional systems\cite{Schulenburg2002}. Our results are the semiclassical description of the exactly factorized  ground state for general spin $S$, as has been done in some frustrated quantum spin chains\cite{Plat-et-al,Acevedo2020}.
As a final remark we can mention a rough estimation for the value of $J_1=J_x$ that corresponds to the end of the dimer phase.
It is easy to see from \eqref{eq:mu} and \eqref{eq:m} that for $J_1=J_x > J_0/3$ the masses $m_3,\mu_3$ and $\mu_4$ become simultaneously negative.
This means that our  non-interacting field theory is unstable with respect of fluctuations $\phi_3$, $\Pi_3$ and $\Pi_4$, and the system should 
be in a ground state described by another theory.

\end{section}

\section{Conclusions}

In this work we use a path integral approach to study the Heisenberg model on the  bilayer honeycomb lattice with in-plane 
first neighbor ($J_1$) and inter-layer first-($J_0$) and second-neighbor ($J_x$) interactions as a model to describe some of the observed features of \bimno. 
Although we have explored different interesting regimes of field theory, we have payed a particular attention to the case  where the values of the magnetic 
couplings are close to those obtained experimentally\cite{Matsuda2019} in \bimno, 
where the dominant coupling is $J_1$. 
In this case, we find that the classical model presents a zero mode parametrized by the relative angle between
the spins of each plane. The low energy effective theory of quantum fluctuations can be written in terms of a symmetric field, related to global 
magnetization, and an antisymmetric field related to the spin imbalance between layers. 
We show that the presence of quantum and thermal fluctuations selects the collinear state from the degenerate manifold of classical ground states. 
This result holds for vorticity free configurations, and coincides with the magnetic order observed experimentally\cite{Matsuda2010}
when a high enough magnetic field is applied, validating the scenario that the experimental observations can be explained with no significant 
frustration in the honeycomb plane but with frustrating intrabilayer interactions and also showing that this mechanism is only present 
in the range of parameters corresponding to the material.
\noindent
If the magnetization of the system is such that $4(S-m) \in \mathds{Z}$ then vortices may proliferate if the renormalized stiffness  
is small enough, driving the system to a gapped phase with only short-range antiferromagnetic ordering. 
This mechanism could explain the short-range correlations observed in \bimno $ \, $at low temperatures and low magnetic fields.\\
Additionally, we show that if $J_0$ is the dominant coupling, the known\cite{Brenig2016} factorized ground state of a valence-bond solid 
composed by a singlet array over the $J_0$ bonds is interpreted in the field theory as a vanishing stiffness and spin wave velocity.
As it was shown for the factorized valence-bond solid ground state, our approach allows also to show that the behavior predicted in the presence of a non-zero magnetization for spin $3/2$ remains true for generic value of the spin. 
We hope that our results will encourage further experimental investigations on \bimno, in particular the ordering transition triggered by an order by disorder mechanism.


\begin{thebibliography}{0}
\expandafter\ifx\csname natexlab\endcsname\relax\def\natexlab#1{#1}\fi
\expandafter\ifx\csname bibnamefont\endcsname\relax
  \def\bibnamefont#1{#1}\fi
\expandafter\ifx\csname bibfnamefont\endcsname\relax
  \def\bibfnamefont#1{#1}\fi
\expandafter\ifx\csname citenamefont\endcsname\relax
  \def\citenamefont#1{#1}\fi
\expandafter\ifx\csname url\endcsname\relax
  \def\url#1{\texttt{#1}}\fi
\expandafter\ifx\csname urlprefix\endcsname\relax\def\urlprefix{URL }\fi
\providecommand{\bibinfo}[2]{#2}
\providecommand{\eprint}[2][]{\url{#2}}

\end{thebibliography}


\begin{thebibliography}{99}
  \expandafter\ifx\csname natexlab\endcsname\relax\def\natexlab#1{#1}\fi
  \expandafter\ifx\csname bibnamefont\endcsname\relax
    \def\bibnamefont#1{#1}\fi
  \expandafter\ifx\csname bibfnamefont\endcsname\relax
    \def\bibfnamefont#1{#1}\fi
  \expandafter\ifx\csname citenamefont\endcsname\relax
    \def\citenamefont#1{#1}\fi
  \expandafter\ifx\csname url\endcsname\relax
    \def\url#1{\texttt{#1}}\fi
  \expandafter\ifx\csname urlprefix\endcsname\relax\def\urlprefix{URL }\fi
  \providecommand{\bibinfo}[2]{#2}
  \providecommand{\eprint}[2][]{\url{#2}}
 
 
 
 \bibitem{Moessner2006}
 R. Moessner and A.P. Ramirez,
 Phys. Today {\bf 59}, 24 (2006).
 
 \bibitem{Balents2010}
 L. Balents, Nature {\bf 464}, 199 (2010).
 
 \bibitem{Wen1990}
 X.-G. Wen,
 Int. J. Mod. Phys. B, {\bf 04}, 239 (1990).
 
 \bibitem{Wen2002}
 X.-G. Wen,
 Phys. Rev. B {\bf 65}, 165113 (2002).
 
 \bibitem{Kitaev2003}
 A. Kitaev, Ann. Phys. (N.Y.) {\bf 303}, 2 (2003).
 
 \bibitem{Benjamin2015}
 S. Benjamin and J. Kelly,
 Nat. Mater. {\bf 14}, 561 (2015).
 
 \bibitem{Riste2015}
 D. Rist\`e, S. Poletto, M.-Z. Huang, A. Bruno, V. Vesterinen,
 O.-P. Saira, and L. DiCarlo, Nat. Commun. {\bf 6}, 7983 (2015).
 
 \bibitem{Corcoles2015}
 A. D. C\'orcoles, E. Magesan, S. J. Srinivasan, A. W. Cross, M. Steffen,
 J. M. Gambetta, and J. M. Chow, Nat. Commun. {\bf 6}, 7979 (2015).
 
\bibitem[{\citenamefont{Mulder et~al.}(2010)\citenamefont{Mulder, Ganesh,
   Capriotti, and Paramekanti}}]{Mulder}
 \bibinfo{author}{\bibfnamefont{A.}~\bibnamefont{Mulder}},
   \bibinfo{author}{\bibfnamefont{R.}~\bibnamefont{Ganesh}},
   \bibinfo{author}{\bibfnamefont{L.}~\bibnamefont{Capriotti}},
   \bibnamefont{and}
   \bibinfo{author}{\bibfnamefont{A.}~\bibnamefont{Paramekanti}},
   \bibinfo{journal}{Phys. Rev. B} \textbf{\bibinfo{volume}{81}},
   \bibinfo{pages}{214419} (\bibinfo{year}{2010}).
 
 \bibitem[{\citenamefont{Okumura et~al.}(2010)\citenamefont{Okumura, Kawamura,
   Okubo, and Motome}}]{Okumura}
 \bibinfo{author}{\bibfnamefont{S.}~\bibnamefont{Okumura}},
   \bibinfo{author}{\bibfnamefont{H.}~\bibnamefont{Kawamura}},
   \bibinfo{author}{\bibfnamefont{T.}~\bibnamefont{Okubo}}, \bibnamefont{and}
   \bibinfo{author}{\bibfnamefont{Y.}~\bibnamefont{Motome}},
   \bibinfo{journal}{J.\ Phys.\ Soc.\ Jpn.} \textbf{\bibinfo{volume}{79}},
   \bibinfo{pages}{114705} (\bibinfo{year}{2010}).
 
 \bibitem[{\citenamefont{Wang}(2010)}]{Wang}
 \bibinfo{author}{\bibfnamefont{F.}~\bibnamefont{Wang}}, \bibinfo{journal}{Phys.
   Rev. B} \textbf{\bibinfo{volume}{82}}, \bibinfo{pages}{024419}
   (\bibinfo{year}{2010}).
 
 \bibitem[{\citenamefont{Mosadeq et~al.}(2011)\citenamefont{Mosadeq, Shahbazi,
   and Jafari}}]{Mosadeq}
 \bibinfo{author}{\bibfnamefont{H.}~\bibnamefont{Mosadeq}},
   \bibinfo{author}{\bibfnamefont{F.}~\bibnamefont{Shahbazi}}, \bibnamefont{and}
   \bibinfo{author}{\bibfnamefont{S.}~\bibnamefont{Jafari}},
   \bibinfo{journal}{Journal of Physics: Condensed Matter}
   \textbf{\bibinfo{volume}{23}}, \bibinfo{pages}{226006}
   (\bibinfo{year}{2011}).
 
 \bibitem[{\citenamefont{Cabra et~al.}(2011{\natexlab{a}})\citenamefont{Cabra,
   Lamas, and Rosales}}]{Cabra_honeycomb_prb}
 \bibinfo{author}{\bibfnamefont{D.~C.} \bibnamefont{Cabra}},
   \bibinfo{author}{\bibfnamefont{C.~A.} \bibnamefont{Lamas}}, \bibnamefont{and}
   \bibinfo{author}{\bibfnamefont{H.~D.} \bibnamefont{Rosales}},
   \bibinfo{journal}{Phys. Rev. B} \textbf{\bibinfo{volume}{83}},
   \bibinfo{pages}{094506} (\bibinfo{year}{2011}{\natexlab{a}}).
   
 \bibitem[{\citenamefont{Oshikawa et~al.}(1997)\citenamefont{Oshikawa, Masaki and Yamanaka, Masanori and Affleck, Ian}}]{OYA}
 \bibinfo{author}{\bibfnamefont{Masaki} \bibnamefont{Oshikawa}},
   \bibinfo{author}{\bibfnamefont{Masanori} \bibnamefont{Yamanaka}}, \bibnamefont{and}
   \bibinfo{author}{\bibfnamefont{Ian} \bibnamefont{Affleck}},
   \bibinfo{journal}{Phys. Rev. Lett.} \textbf{\bibinfo{volume}{78}},
   \bibinfo{pages}{1984--1987} (\bibinfo{year}{1997}).
   
   

 
 \bibitem[{\citenamefont{Ganesh et~al.}(2011{\natexlab{a}})\citenamefont{Ganesh,
   Sheng, Kim, and Paramekanti}}]{Ganesh_2011}
 \bibinfo{author}{\bibfnamefont{R.}~\bibnamefont{Ganesh}},
   \bibinfo{author}{\bibfnamefont{D.}~\bibnamefont{Sheng}},
   \bibinfo{author}{\bibfnamefont{Y.-J.} \bibnamefont{Kim}}, \bibnamefont{and}
   \bibinfo{author}{\bibfnamefont{A.}~\bibnamefont{Paramekanti}},
   \bibinfo{journal}{Phys.\ Rev.\ B} \textbf{\bibinfo{volume}{83}},
   \bibinfo{pages}{144414} (\bibinfo{year}{2011}{\natexlab{a}}).
 
 \bibitem[{\citenamefont{Albuquerque et~al.}(2011)\citenamefont{Albuquerque,
   Schwandt, Het\'{e}nyi, Capponi, Mambrini, and L\"auchli}}]{Albuquerque}
 \bibinfo{author}{\bibfnamefont{A.~F.} \bibnamefont{Albuquerque}},
   \bibinfo{author}{\bibfnamefont{D.}~\bibnamefont{Schwandt}},
   \bibinfo{author}{\bibfnamefont{B.}~\bibnamefont{Het\'{e}nyi}},
   \bibinfo{author}{\bibfnamefont{S.}~\bibnamefont{Capponi}},
   \bibinfo{author}{\bibfnamefont{M.}~\bibnamefont{Mambrini}}, \bibnamefont{and}
   \bibinfo{author}{\bibfnamefont{A.~M.} \bibnamefont{L\"auchli}},
   \bibinfo{journal}{Phys. Rev. B} \textbf{\bibinfo{volume}{84}},
   \bibinfo{pages}{024406} (\bibinfo{year}{2011}).
 
 \bibitem[{\citenamefont{Clark et~al.}(2011)\citenamefont{Clark, Abanin, and
   Sondhi}}]{Clark}
 \bibinfo{author}{\bibfnamefont{B.}~\bibnamefont{Clark}},
   \bibinfo{author}{\bibfnamefont{D.}~\bibnamefont{Abanin}}, \bibnamefont{and}
   \bibinfo{author}{\bibfnamefont{S.}~\bibnamefont{Sondhi}},
   \bibinfo{journal}{Phys. Rev. Lett.} \textbf{\bibinfo{volume}{107}},
   \bibinfo{pages}{087204} (\bibinfo{year}{2011}).
 
 \bibitem[{\citenamefont{Cabra et~al.}(2011{\natexlab{b}})\citenamefont{Cabra,
   Lamas, and Rosales}}]{Cabra_honeycomb_2}
 \bibinfo{author}{\bibfnamefont{D.}~\bibnamefont{Cabra}},
   \bibinfo{author}{\bibfnamefont{C.}~\bibnamefont{Lamas}}, \bibnamefont{and}
   \bibinfo{author}{\bibfnamefont{H.}~\bibnamefont{Rosales}},
   \bibinfo{journal}{Mod. Phys. Lett. B} \textbf{\bibinfo{volume}{25}},
   \bibinfo{pages}{891} (\bibinfo{year}{2011}{\natexlab{b}}).
 
 \bibitem[{\citenamefont{Mezzacapo and Boninsegni}(2012)}]{Mezzacapo}
 \bibinfo{author}{\bibfnamefont{F.}~\bibnamefont{Mezzacapo}} \bibnamefont{and}
   \bibinfo{author}{\bibfnamefont{M.}~\bibnamefont{Boninsegni}},
   \bibinfo{journal}{Phys. Rev. B} \textbf{\bibinfo{volume}{85}},
   \bibinfo{pages}{060402(R)} (\bibinfo{year}{2012}).
 
 \bibitem[{\citenamefont{Bishop et~al.}(2012)\citenamefont{Bishop, Li., Farnell,
   and Campbell}}]{Bishop_2012}
 \bibinfo{author}{\bibfnamefont{R.~F.} \bibnamefont{Bishop}},
   \bibinfo{author}{\bibfnamefont{P.~H.~Y.} \bibnamefont{Li.}},
   \bibinfo{author}{\bibfnamefont{D.~J.~J.} \bibnamefont{Farnell}},
   \bibnamefont{and} \bibinfo{author}{\bibfnamefont{C.~E.}
   \bibnamefont{Campbell}}, \bibinfo{journal}{J.\ Phys.: Condens.\ Matter}
   \textbf{\bibinfo{volume}{24}}, \bibinfo{pages}{236002}
   (\bibinfo{year}{2012}).
 
 \bibitem[{\citenamefont{Li et~al.}(2012)\citenamefont{Li, Bishop, Farnell, and
   Campbell}}]{Li_2012_honeyJ1-J2-J3}
 \bibinfo{author}{\bibfnamefont{P.~H.~Y.} \bibnamefont{Li}},
   \bibinfo{author}{\bibfnamefont{R.~F.} \bibnamefont{Bishop}},
   \bibinfo{author}{\bibfnamefont{D.~J.~J.} \bibnamefont{Farnell}},
   \bibnamefont{and} \bibinfo{author}{\bibfnamefont{C.~E.}
   \bibnamefont{Campbell}}, \bibinfo{journal}{Phys.\ Rev.\ B}
   \textbf{\bibinfo{volume}{86}}, \bibinfo{pages}{144404}
   (\bibinfo{year}{2012}).
 
 \bibitem[{\citenamefont{Bishop et~al.}(2013)\citenamefont{Bishop, Li, Farnell,
   and Campbell}}]{Bishop_2013}
 \bibinfo{author}{\bibfnamefont{R.~F.} \bibnamefont{Bishop}},
   \bibinfo{author}{\bibfnamefont{P.~H.~Y.} \bibnamefont{Li}},
   \bibinfo{author}{\bibfnamefont{D.~J.~J.} \bibnamefont{Farnell}},
   \bibnamefont{and} \bibinfo{author}{\bibfnamefont{C.~E.}
   \bibnamefont{Campbell}}, \bibinfo{journal}{J. Phys.: Condens. Matter}
   \textbf{\bibinfo{volume}{25}}, \bibinfo{pages}{306002}
   (\bibinfo{year}{2013}).
 
 \bibitem[{\citenamefont{Gong et~al.}(2013)\citenamefont{Gong, Sheng, Motrunich,
   and Fisher}}]{Fisher_2013}
 \bibinfo{author}{\bibfnamefont{S.-S.} \bibnamefont{Gong}},
   \bibinfo{author}{\bibfnamefont{D.}~\bibnamefont{Sheng}},
   \bibinfo{author}{\bibfnamefont{O.~I.} \bibnamefont{Motrunich}},
   \bibnamefont{and} \bibinfo{author}{\bibfnamefont{M.~P.}
   \bibnamefont{Fisher}}, \bibinfo{journal}{Phys. Rev. B}
   \textbf{\bibinfo{volume}{88}}, \bibinfo{pages}{165138}
   (\bibinfo{year}{2013}).
 
 \bibitem[{\citenamefont{Ganesh et~al.}(2013)\citenamefont{Ganesh, van~den
   Brink, and Nishimoto}}]{Ganesh_PRL_2013}
 \bibinfo{author}{\bibfnamefont{R.}~\bibnamefont{Ganesh}},
   \bibinfo{author}{\bibfnamefont{J.}~\bibnamefont{van~den Brink}},
   \bibnamefont{and}
   \bibinfo{author}{\bibfnamefont{S.}~\bibnamefont{Nishimoto}},
   \bibinfo{journal}{Phys. Rev. Lett.} \textbf{\bibinfo{volume}{110}},
   \bibinfo{pages}{127203} (\bibinfo{year}{2013}).
 
 \bibitem[{\citenamefont{Zhu et~al.}(2013)\citenamefont{Zhu, Huse, and
   White}}]{Zhu_PRL_2013}
 \bibinfo{author}{\bibfnamefont{Z.}~\bibnamefont{Zhu}},
   \bibinfo{author}{\bibfnamefont{D.~A.} \bibnamefont{Huse}}, \bibnamefont{and}
   \bibinfo{author}{\bibfnamefont{S.~R.} \bibnamefont{White}},
   \bibinfo{journal}{Phys. Rev. Lett.} \textbf{\bibinfo{volume}{110}},
   \bibinfo{pages}{127205} (\bibinfo{year}{2013}).
 
 \bibitem[{\citenamefont{Zhang and Lamas}(2013)}]{Zhang_PRB_2013}
 \bibinfo{author}{\bibfnamefont{H.}~\bibnamefont{Zhang}} \bibnamefont{and}
   \bibinfo{author}{\bibfnamefont{C.}~\bibnamefont{Lamas}},
   \bibinfo{journal}{Phys. Rev. B} \textbf{\bibinfo{volume}{87}},
   \bibinfo{pages}{024415} (\bibinfo{year}{2013}).
 
 
 \bibitem[{\citenamefont{Beca}(2013)}]{Beca_2017}
 \bibinfo{author}{\bibfnamefont{F.}~\bibnamefont{Ferrari}},
 \bibinfo{author}{\bibfnamefont{S.}~\bibnamefont{Bieri}} \bibnamefont{and}
   \bibinfo{author}{\bibfnamefont{F.}~\bibnamefont{Becca}},
   \bibinfo{journal}{Phys. Rev. B} \textbf{\bibinfo{volume}{96}},
   \bibinfo{pages}{104401} (\bibinfo{year}{2017}).

 \bibitem[{\citenamefont{Smirnova et~al.}(2009)\citenamefont{Smirnova, Azuma,
   Kumada, Kusano, Matsuda, Shimakawa, Takei, Yonesaki, and
   Kinomura}}]{smirnova2009synthesis}
 \bibinfo{author}{\bibfnamefont{O.}~\bibnamefont{Smirnova}},
   \bibinfo{author}{\bibfnamefont{M.}~\bibnamefont{Azuma}},
   \bibinfo{author}{\bibfnamefont{N.}~\bibnamefont{Kumada}},
   \bibinfo{author}{\bibfnamefont{Y.}~\bibnamefont{Kusano}},
   \bibinfo{author}{\bibfnamefont{M.}~\bibnamefont{Matsuda}},
   \bibinfo{author}{\bibfnamefont{Y.}~\bibnamefont{Shimakawa}},
   \bibinfo{author}{\bibfnamefont{T.}~\bibnamefont{Takei}},
   \bibinfo{author}{\bibfnamefont{Y.}~\bibnamefont{Yonesaki}}, \bibnamefont{and}
   \bibinfo{author}{\bibfnamefont{N.}~\bibnamefont{Kinomura}},
   \bibinfo{journal}{Journal of the American Chemical Society}
   \textbf{\bibinfo{volume}{131}}, \bibinfo{pages}{8313} (\bibinfo{year}{2009}).


 \bibitem[{\citenamefont{Ganesh et~al.}(2011{\natexlab{b}})\citenamefont{Ganesh,
   Isakov, and Paramekanti}}]{Ganesh_QMC}
 \bibinfo{author}{\bibfnamefont{R.}~\bibnamefont{Ganesh}},
   \bibinfo{author}{\bibfnamefont{S.~V.} \bibnamefont{Isakov}},
   \bibnamefont{and}
   \bibinfo{author}{\bibfnamefont{A.}~\bibnamefont{Paramekanti}},
   \bibinfo{journal}{Phys. Rev. B} \textbf{\bibinfo{volume}{84}},
   \bibinfo{pages}{214412} (\bibinfo{year}{2011}{\natexlab{b}}).
 
 \bibitem[{\citenamefont{Oitmaa and Singh}(2012)}]{Oitmaa_2012}
 \bibinfo{author}{\bibfnamefont{J.}~\bibnamefont{Oitmaa}} \bibnamefont{and}
   \bibinfo{author}{\bibfnamefont{R.}~\bibnamefont{Singh}},
   \bibinfo{journal}{Phys. Rev. B} \textbf{\bibinfo{volume}{85}},
   \bibinfo{pages}{014428} (\bibinfo{year}{2012}).
 
 \bibitem[{\citenamefont{Zhang et~al.}(2014)\citenamefont{Zhang, Arlego, and
   Lamas}}]{Zhang2014}
 \bibinfo{author}{\bibfnamefont{H.}~\bibnamefont{Zhang}},
   \bibinfo{author}{\bibfnamefont{M.}~\bibnamefont{Arlego}}, \bibnamefont{and}
   \bibinfo{author}{\bibfnamefont{C.~A.} \bibnamefont{Lamas}},
   \bibinfo{journal}{Phys. Rev. B} \textbf{\bibinfo{volume}{89}},
   \bibinfo{pages}{024403} (\bibinfo{year}{2014}).
 
 \bibitem[{\citenamefont{Arlego et~al.}(2014)\citenamefont{Arlego, Lamas, and
   Zhang}}]{Arlego201415}
 \bibinfo{author}{\bibfnamefont{M.}~\bibnamefont{Arlego}},
   \bibinfo{author}{\bibfnamefont{C.~A.} \bibnamefont{Lamas}}, \bibnamefont{and}
   \bibinfo{author}{\bibfnamefont{H.}~\bibnamefont{Zhang}}, \bibinfo{journal}{J.
   Phys.: Conf. Ser.} \textbf{\bibinfo{volume}{568}}, \bibinfo{pages}{042019}
   (\bibinfo{year}{2014}).
 
 \bibitem[{\citenamefont{Zhang et~al.}(2016)\citenamefont{Zhang, Lamas, Arlego,
   and Brenig}}]{Brenig2016}
 \bibinfo{author}{\bibfnamefont{H.}~\bibnamefont{Zhang}},
   \bibinfo{author}{\bibfnamefont{C.~A.} \bibnamefont{Lamas}},
   \bibinfo{author}{\bibfnamefont{M.}~\bibnamefont{Arlego}}, \bibnamefont{and}
   \bibinfo{author}{\bibfnamefont{W.}~\bibnamefont{Brenig}},
   \bibinfo{journal}{Phys. Rev. B} \textbf{\bibinfo{volume}{93}},
   \bibinfo{pages}{235150} (\bibinfo{year}{2016}).
 
 \bibitem[{\citenamefont{Bishop and Li}(2017)}]{bishop2017frustrated}
 \bibinfo{author}{\bibfnamefont{R.~F.} \bibnamefont{Bishop}} \bibnamefont{and}
   \bibinfo{author}{\bibfnamefont{P.~H.~Y.} \bibnamefont{Li}},
   \bibinfo{journal}{Phys. Rev. B} \textbf{\bibinfo{volume}{95}},
   \bibinfo{pages}{134414} (\bibinfo{year}{2017}).
 
 \bibitem[{\citenamefont{Krokhmalskii et~al.}(2017)\citenamefont{Krokhmalskii,
   Baliha, Derzhko, Schulenburg, and Richter}}]{Richter2017}
 \bibinfo{author}{\bibfnamefont{T.}~\bibnamefont{Krokhmalskii}},
   \bibinfo{author}{\bibfnamefont{V.}~\bibnamefont{Baliha}},
   \bibinfo{author}{\bibfnamefont{O.}~\bibnamefont{Derzhko}},
   \bibinfo{author}{\bibfnamefont{J.}~\bibnamefont{Schulenburg}},
   \bibnamefont{and} \bibinfo{author}{\bibfnamefont{J.}~\bibnamefont{Richter}},
   \bibinfo{journal}{Phys. Rev. B} \textbf{\bibinfo{volume}{95}},
   \bibinfo{pages}{094419} (\bibinfo{year}{2017}).

\bibitem{Matsuda2010}
Matsuda et al., Phys. Rev. B, 105,187201 (2010).

\bibitem[{\citenamefont{Mastuda et~al.}(2019)\citenamefont{Matsuda, Dissanayake,
   Abernathy, Qiu, Copley, Kumada, and Azuma.}}]{Matsuda2019}
 \bibinfo{author}{\bibfnamefont{M.}~\bibnamefont{Matsuda}},
   \bibinfo{author}{\bibfnamefont{S. E.}~\bibnamefont{Dissanayake}},
   \bibinfo{author}{\bibfnamefont{D. L.}~\bibnamefont{Abernathy}},
      \bibinfo{author}{\bibfnamefont{Y.}~\bibnamefont{Qiu}},
   \bibinfo{author}{\bibfnamefont{ J. R. D.}~\bibnamefont{Copley}},
      \bibinfo{author}{\bibfnamefont{N.}~\bibnamefont{Kumada}},
   \bibnamefont{and}
   \bibinfo{author}{\bibfnamefont{M.}~\bibnamefont{Azuma}},
   \bibinfo{journal}{Phys. Rev. B} \textbf{\bibinfo{volume}{100}},
   \bibinfo{pages}{134430} (\bibinfo{year}{2019}).

  
 
%
%
%
%
%
  \bibitem{Alaei2017} 
   \bibinfo{author}{\bibfnamefont{TM}~\bibnamefont{Alaei et al.}}, 
   \bibinfo{journal}{Phys. Rev. B} \textbf{\bibinfo{volume}{96}},
   \bibinfo{pages}{140404} (\bibinfo{year}{2017}).
  

 \bibitem{Haldane} 
   \bibinfo{author}{\bibfnamefont{F. D. M.}~\bibnamefont{Haldane}}, 
   \bibinfo{journal}{Phys. Rev. Lett.} \textbf{\bibinfo{volume}{57}},
   \bibinfo{pages}{1488} (\bibinfo{year}{1986}).
   
   \bibitem{TTH} 
   \bibinfo{author}{\bibfnamefont{A.}~\bibnamefont{Tanaka}}, 
    \bibinfo{author}{\bibfnamefont{K.}~\bibnamefont{Totsuka}}, 
     \bibinfo{author}{\bibfnamefont{X.}~\bibnamefont{Hu}}, 
   \bibinfo{journal}{Phys. Rev. B} \textbf{\bibinfo{volume}{79}},
   \bibinfo{pages}{064412} (\bibinfo{year}{2009}).
   
   
\bibitem{LCPR} C. A. Lamas, D. C. Cabra, P. Pujol, and G. L. Rossini, Eur. Phys. J. B, 88, 176 (2015). 

\bibitem{Lamas2011} C. A. Lamas, S. Capponi, P. Pujol, Phys. Rev. B, 84, 115125 (2011).

 \bibitem[{\citenamefont{G\'omez Albarrac\'in et al}(2016)\citenamefont{G\'omez Albarrac\'in and Rosales}}]{FlaviaDiegoMC}
 \bibinfo{author}{\bibfnamefont{F.}~\bibnamefont{G\'omez Albarrac\'in}},
\bibnamefont{and}
   \bibinfo{author}{\bibfnamefont{D.}~\bibnamefont{Rosales}},
   \bibinfo{journal}{Phys. Rev. B} \textbf{\bibinfo{volume}{93}},
   \bibinfo{pages}{144413} (\bibinfo{year}{2016}).

\bibitem{Schulenburg2002} Schulenburg et al. Phys. Rev. Lett. 88, 167207 (2002)

\bibitem{Plat-et-al} Plat, X. and Capponi, S. and Pujol, P., Phys. Rev. B 85,174423 (2012).

\bibitem{Acevedo2020} S. Acevedo, P. Pujol, C. A. Lamas, Phys. Rev. B 102, 195139 (2020)



%
%
%
%





%
%
%
%
%
%
%
%
%
%
%
%
%
%
%
%
%
















   
 
%
%
%



\end{thebibliography}
\end{document}